\documentclass[10pt,letterpaper,twocolumn]{article} 

\usepackage{ol2}
\usepackage[draft]{hyperref}
\usepackage{amsmath,amssymb,dcolumn,bm,graphicx}
\begin{document}

\twocolumn[

\title{Probing non-Abelian anyonic statistics with cold atoms\\ in an optical lattice}

\author{Sheng Liu and Zheng-Yuan Xue*}

\address {Laboratory of Quantum Information Technology, and
School of Physics and Telecommunication Engineering,\\
 South China Normal University, Guangzhou 510006, China\\
$^*$Corresponding author: zyxue@scnu.edu.cn}

\begin{abstract}
We propose a scheme to probe the non-Abelian statistics of the collective anyonic excitation in Kitaev's honeycomb model
with cold atoms in an optical lattice. The generation of the anyonic excitation can be realized by simple rotating
operation acting on an effective spin-1/2 system, which is encoded in the atomic hyperfine energy levels.
The non-Abelian nature  of the anyonic excitation is manifested by the  braiding of four vortices,
which  leads to different operations on the subspace  of degenerate ground states, and thus results in different final states.
Here, by introducing an ancilla atom, the effective control over the lattice atoms can be realized and the final  different states
can also be imprinted on the ancilla and further distinguished by measurement.

\end{abstract}

\ocis{000.1600, 
140.3325, 
270.5585 
}
]

\noindent
Quantum statistics is one of the basic concepts in quantum physics \cite{Dirac}.
It's also the root of some condensed matter physical phenomena, such as, superfluid, Bose-Einstein condensates, etc.
In three-dimension, due to the rotating and time reversal symmetries, particles
obey either the Bose-Einstein or Fermi-Dirac statistics corresponding to their statistical
phase $\alpha$ $(\alpha=\theta/\pi)$ to be $0$ or $1$. But, there exist cases with $\alpha\neq0$ or $1$ corresponding to more
exotic particles called anyons, which maybe  either Abelian or non-Abelian because of symmetry breaking
in 2-dimension \cite{Khare}. The elementary excitation with fractional charge in fractional
quantum Hall effects (FQHE) is anyon satisfying exotic statistics \cite{Laughlin,l2}, which
provides a good platform to understand anyonic statistics. For Abelian anyons, the anyonic wave-function
acquires a phase factor coming from the winding of two quasiparticles, which is just the one-dimensional representation of
the braid group $B_N$ (for $N$ indistinguishable particles). For non-Abelian anyons, quasiparticles
exchange would be represented by a matrix acting on the subspace of the degenerate ground states, which is
an exactly noncommutative representation of the braid group in higher dimension \cite{Moore}.
In particular, Kitaev  showed \cite{Kitaev1} that non-Abelian anyon is a  promising candidate
for topological quantum computation, which is fault-tolerant at the hardware level.

Up to now, no experiment has unambiguously verified the existence of Abelian or
non-Abelian anyons although promising progress have been made in  $\nu=5/2$
FQHE state \cite{Stern}. This is  mainly due to the extremely complicated noise
channels in a quantum many-body solid-state system. On the other hand, a cold
atom system possesses unprecedented possibility of controlling almost all
relevant physical parameters \cite{bloch} and thus been recognized as an ideal
system for quantum simulation. Proposals of observing the non-Abelian
statistics are also presented with cold atoms in \emph{s}-wave \cite{Zhu,xue2}
and \emph{p}-wave \cite{Tewari} superfluidity. For Kitaev's toric model, it
has been proposed  to simulate the anyonic interferometry in virous systems
\cite{han,xue,cirac,jiang,zhang,you,v1,v2} with experimental verification in optical systems
\cite{zhangj,lucy,pachos} using a method of generating dynamically the ground
states and the excitations of the model Hamiltonian. However,  there is few
reports about the verification of non-Abelian statistics in the Kitaev
honeycomb lattice model \cite{Kitaev2}. Recently, theoretical implementation of
this model has been proposed with cold atoms in an honeycomb lattice \cite{Duan}.
Meanwhile, elementary experimental verification of the
Dirac physics associated has also been made with cold atoms in a honeycomb lattice \cite{h1}
based on theoretical proposals in  \cite{h2,h3}.

Here, motivated by the above advances, we propose a concrete example of probing  the exotic non-Abelian
statistics in the Kitaev honeycomb model with cold atoms in optical lattice.
The honeycomb lattice is constructed from cold atoms located in an optical lattice and
the anisotropic spin interaction is induced from spin-dependent tunneling between the nearest neighbor of the lattice \cite{Duan}.
To get effective control over the lattice atoms, following Ref. \cite{cirac}, we introduce an auxiliary  atom.
Then, creation and manipulation of anyons are implemented by Rydberg gates based on dipole-dipole
interaction between the ancilla  and lattice atoms. Finally, the exotic non-Abelian statistics
can be recorded by the ancilla during the implemented braiding process.
Therefore, detect the final state of the ancilla can  reveal the non-Abelian statistics.
Furthermore, it is usually difficult to read out the states of the anyons due to the fact that they are degenerated and neutral.
Therefore, we introduce an auxiliary  atom to imprint the difference of the
anyonic final states after braiding. Meanwhile, for probing the non-Abelian
statistics, one needs at least four anyons. If only two anyons are involved,
the final state after braiding  will be different from the initial state by a phase factor.
However, this difference is insufficient to verify that the involved anyons are of
non-Abelian nature since braiding of Abelian anyons may also cause such a phase difference.

\begin{figure}[tbp]
\begin{center}
\includegraphics[width=8.5cm,]{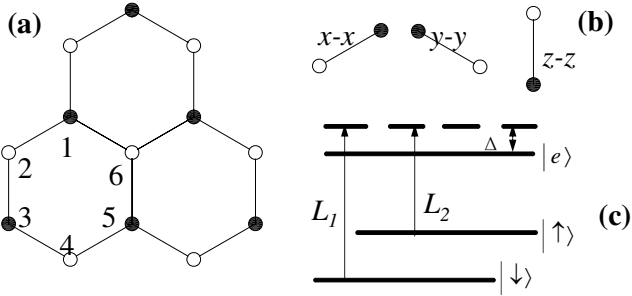}
\end{center}
\caption{Implementation of the Kitaev model.
(a) The model is a honeycomb lattice model of spin 1/2 system in a plane.
The index of the 6 atom within a plaquette is indicated.
(b) There are 3 different interaction between the nearest neighbor of the lattice
along the $x-x$, $y-y$ and $z-z$ directions.
(c) The  pseudo spin states are encoded by $|\uparrow\rangle$ and
$|\downarrow\rangle$ of an atom,  an excited state $|e\rangle$ is also introduced.
Two coupling laser beams $L_1$ and $L_2$ are used
to realize the interaction in three independent directions as we needed.}
\end{figure}

Let us begin with a brief review of the  Kitaev honeycomb  lattice model,
which is an anisotropic spin-1/2 model with different nearest neighbor interactions in different bonds \cite{Kitaev2}
\begin{equation}
\label{K}
H=-J_x\sum_{x-links}\sigma_j^x\sigma_k^x-J_y \sum_{y-links}\sigma_j^y\sigma_k^y-J_z \sum_{z-links}\sigma_j^z\sigma_k^z,
\end{equation}
where $\sigma_j^\nu$ with $\nu \in \{x, y, z\}$ being the pauli matrix on the site $j$,
$J_\nu$ are coupling strength for different links as indicated in Fig. 1.
To get the ground states, diagonalization of the model $H$ is needed, which can be greatly simplified
by constructing a set of integrals of motion $W_p=\sigma_1^x\sigma_2^y\sigma_3^z\sigma_4^x\sigma_5^y\sigma_6^z$
with $p$ being a label of a plaquette of the lattice and the spin index is labeled as in Fig. 1(a).
Expanding every spin into a four Majorana operators notation,  the model Hamiltonian in the enlarged space reads \cite{Kitaev2}
\begin{eqnarray}
\label{E}
H=\displaystyle\frac{i}{4}\sum_{j,k}\hat{A}_{jk}c_jc_k,
\end{eqnarray}
where $\hat{A}_{jk}= 2J_\alpha\hat{u}_{jk}$ if spins $j$ and $k$ are connected,
otherwise $\hat{A}_{jk}= 0$. Here, $\hat{u}_{jk}= ib_j^\alpha b_k^\alpha$
with $b_j$ and $c_j$ being Majorana operators, $[H, \hat{u}_{jk}]=0$,
$\hat{u}_{jk}^\dagger=\hat{u}_{jk}$, and $\hat{u}_{jk}^2=1$.
We first divide the total original Hilbert space into different sector which is eigenspace of $W_p$.
So $W_p$ can be transformed to the eigenspaces of a set of  states given by a certain
configuration of eigenvalue of  $\hat{u}_{jk}$. The connection between $w_p$ (eigenvalue of the $W_p$ )
and $u_{jk}$ (eigenvalue of $\hat{u}_{jk}$) is  $w_p=\prod_{(j,k)\in boundary(p)}u_{jk}$.
The variables $u_{jk}$ and the numbers $w_p$ can be interpreted as a $\mathbb{Z}_2$ gauge field and
the magnetic flux through the plaquette $p$, respectively. We say that the plaquette $p$ carries
a vortex when $w_p=-1$ and the ground states are denoted by the vortex-free sector,
i.e. all  $w_p=1$. Then go to the momentum representation via the Fourier transformation,
we can get the ground energy, ground states and the phase diagram.
In the triangle region of $|J_x|\leq|J_y|+|J_z|$, $|J_y|\leq|J_z|+|J_x|$ and  $|J_z|\leq|J_x|+|J_y|$,
it's gapless phase B with  non-Abelian anyonic excitation \cite{Kitaev2};
otherwise, it's gapped phase A with Abelian anyonic excitation.
A magnetic field $V=-\sum_j(h_x\sigma_j^x+h_y\sigma_j^y+h_z\sigma_j^z)$
is introduced to open a gap between the ground and the excited states in phase B \cite{Kitaev2}.
This gap will exponentially suppress the interaction between vortices and fermionic modes
near the singularity of the spectrum, and thus makes the statistics of non-Abelian anyon  well-defined.

\begin{figure*}
\begin{center}
\includegraphics[width=12cm,]{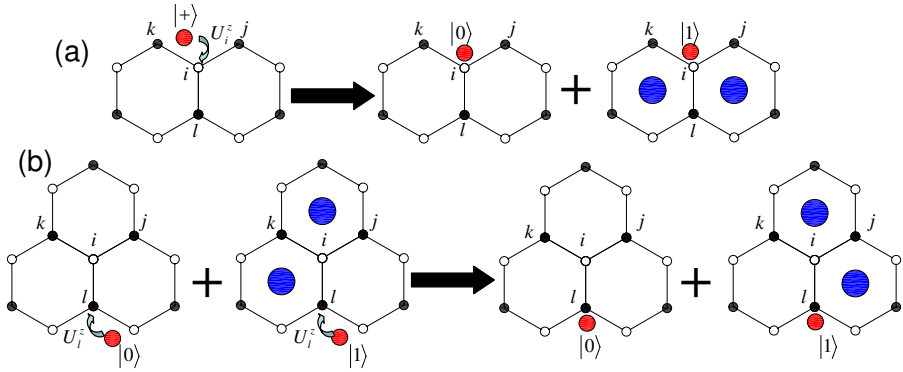}
\end{center}
\caption{(color online) The creation and manipulation of non-Abelian
anyons. (a) Initial state of the ancilla atom (the red dot) is
$|+\rangle_a\propto|0\rangle_a+|1\rangle_a$ and the solid blue
circle denote a vortex excitation bound to a plaquette. When moving
the ancilla  close enough to the code atom, 2-qubit $U_i^z$ gate is
achieved. (b) The manipulation of the anyons, only when the ancilla
is in state $|1\rangle_a$, an effective operation will act on the
atom transforming the $w_p$ from one to the other sector, and thus
the vortex configuration can be changed as we needed.}
\end{figure*}

Now, we turn to construct the  honeycomb model of Kitaev using ultracold fermionic atoms in an optical lattice.
Following Ref. \cite{Duan}, we consider a cloud of ultracold  Rb$^{87}$  atoms trapped in an optical lattice
which are formed by interfering standing laser beams. First, completely suppress the tunneling and
the spin interchange in the vertical $Z$ direction by raising the corresponding barrier potential,
and thus the atoms distribution form a two-dimensional configuration in the $X-Y$ plane.
Assuming that the system in the Mott insulator regime and the atomic density is roughly one atom per site.
Each atom is treated as a two-level system, labeled with the effective spin index $\sigma=\uparrow,\downarrow$,
with an energy splitting of 6834 MHz.
To form the honeycomb lattice, we apply trapping potential  $V_j(x,y)=V_0 sin^2[k_\|(x \cos \theta_j+y
 \sin\theta_j)+\phi_0]$ formed by a pair of blue-detuned traveling
laser beams above the X-Y plane with an angle $\phi_\|$, where,
$j=1,2,3$, $\theta_{1(2,3)}=\pi/6 (\pi/2,-\pi/6)$,
$\phi_\|=2\arcsin(1/\sqrt3)$, and $\phi_0=\pi/2$. To engineer the
wanted interaction in different directions, applying the  following
three blue-detuned standing-wave lase beams
$V_{\nu\sigma}(x,y)=V_{\nu\sigma}\sin^2[k(x\cos\theta^\prime_\nu+y\sin\theta^\prime_\nu)]$
in the X-Y plane along certain directions, where    $\nu$ denote the
tunneling direction x-x, y-y, z-z, $\theta_{x-x}^\prime=-\pi$,
$\theta_{y-y}^\prime=\pi$, $\theta_{z-z}^\prime=\pi/3$. These laser
beams will introduce spin-dependent potentials
$V_{\nu\sigma}=V_{\nu+}|+\rangle_\nu\langle+|+V_{\nu-}|-\rangle_\nu\langle-|$,
where $|+\rangle_\nu$ and $|-\rangle_\nu$ are the eigenvectors of
the eigenvalues $1$ or $-1$ of Pauli operator $\sigma^\nu$ (cf. Fig.
1), respectively. That is to say, different spin directions will feel
different effective potential, and thus have different tunneling
rates. The potential depth between the nearest neighbor along the
$\nu$ direction can be changed from $V_0/4$ to $V_0/4+V_{\nu\sigma}$
\cite{Duan}. The parameters $V_{\nu\pm}$ can be tuned by varying the
laser intensity of $L_1$ and $L_2$ in the $\nu$ direction. Here, we
are interested in the regime $t_{\nu+}\gg t_{\nu-}$, i.e., the atom
can virtually tunnel only when it is in the eigenstate of
$|+\rangle_\nu$. Then, the  spin-dependent potentials will lead to
anisotropic Ising exchange interaction for each tunneling direction, and the
total interaction is in the form of \cite{Duan}
\begin{equation}
\label{C}
H_{eff}=-\sum_{\langle i,j\rangle,\nu}J_\nu\sigma_i^\nu\sigma_j^\nu-h_{\nu}\sum_{j\nu}\sigma^\nu_j.
\end{equation}
In the case of $V_{\nu-}/V_{\nu+}\gg1 $ and $U_{+} \approx U_{-} \approx U_{+-}\approx
U$, where $U_\sigma$ and $ U_{+-}$ are the intra- and interspin coupling energy,
$J_\nu=t^2_{+\nu}/(2U)$ and $h_\nu=4t_{\nu+}^2/U$.  Then, we get a Hamiltonian of Eq. (\ref{K}) with
additional effective magnetic field terms $h_{\nu}\sum_j\sigma_j^\nu$, which can be used  to open the gap in phase B.
Therefore, this implementation naturally leads to the effective  magnetic field terms, and thus makes it more easier for realization.

To probe the non-Abelian statistics, we need at least two pairs of non-Abelian anyons due to their superselection rule.
The generation of a pair of anyons can be realized by a simple $\sigma^z$ rotation acting on a single spin,
which is in a vortex-free configuration. But, the generation process will accompany  with fermionic excitations in the mean time.
Fortunately, the unwanted fermionic excitations can be exponentially suppressed by the energy gap opened by the effective magnetic field.
The $\sigma^z$ rotation may be implemented by laser-atom interaction, but single atom addressing is difficult experimentally.
To avoid such difficulty \cite{cirac}, we consider an ancilla atom that is trapped above $X-Y$ plane by
three standing wave laser beams along three direction with its internal Zeeman levels $|0\rangle_a$ and $|1\rangle_a$.
The $\sigma^z$ rotation can be realized by  Rydberg gates based on dipole-dipole interactions between
the ancilla and a lattice atom \cite{g1,g2}, which maximally reduce the addressing difficulty and eliminates the
need of cooling  both the lattice  and   ancilla to the physical ground states in the cold collisions implementation \cite{Jaksch}.
On the other hand,  this single ancilla atom can be moved by trap potential without decoherence \cite{Kuhr}
and this is important to guarantee our implementation. As a result, 2-qubit unitary
operations $U^{\alpha}_i=|0\rangle_a\langle0|\otimes I+|1\rangle_a\langle1|\otimes\sigma^\alpha_i$ acting on site atom $i$
and the ancilla atom  are implemented, which are sufficient to create
and manipulate the anyons \cite{Kitaev2}, as shown in  Fig. 2.   Experimentally,  an arbitrary qubit hopping time is $8\mu s$
with a nearly unit fidelity \cite{Schrader}. On the condition that the  ancilla atom is in the $|1\rangle_a$ state,
$\sigma^\alpha_i$ rotations can be implemented on site $i$ atom. The generation of two vortex configuration can be represented  as
\begin{equation}
H_{2-v}=\sigma^z_i H \sigma^z_i=H+2J_x\sigma^x_i\sigma^x_j+2J_y\sigma^y_i\sigma^y_k,
\end{equation}
where  $\sigma_i^z$ denotes a $\sigma^z$ rotation on site $i$,
$(i,j)$ label a x-link, and $(i,k)$ labels a y-link. meanwhile,  we
can also get two vortices via a $\sigma_i^x$ or $\sigma_i^y$
rotation but with different anyons configuration. This simple
configuration due to $\hat{u}_{il}=-1$ at the site $i$ along the
$z-z$ direction with the neighbor honeycomb lattice $\omega_p=-1$,
and the vortex-free configuration as it gives $\omega_p=1$  for all
plaquettes $p$. The exotic collective excitation are non-Abelian
anyons bound to the vortices, and the non-Abelian anyons are
Majorana fermions \cite{Kitaev2}. At low temperature, the initial
state of the pair of  Majorana fermions  is typically a vacuum state
$|0\rangle$ \cite{Tewari}.

To make our scheme more robust to decoherence and avoid the problem of single atom addressing,
we use the conditional generation of anyons using $U_i^\alpha$. When $\alpha=z$,
the ancilla atom,  vortex-free ground state $|gs\rangle$ and 2-vortex state $|0\rangle$
are entangled as shown in  Fig. 2, i.e. $|0\rangle_a|gs\rangle+|1\rangle_a|0\rangle $.
Therefore, the exotic non-Ableian statistics can be recorded by the ancilla atom,
i.e., we can also use the state of the ancilla atom to manifest the non-Abelian statistics
of anyon. The state of ancilla atom  can be addressed  independently, and thus can be measured by  standard probing technique.

\begin{figure*}
\begin{center}
\includegraphics[width=12cm,]{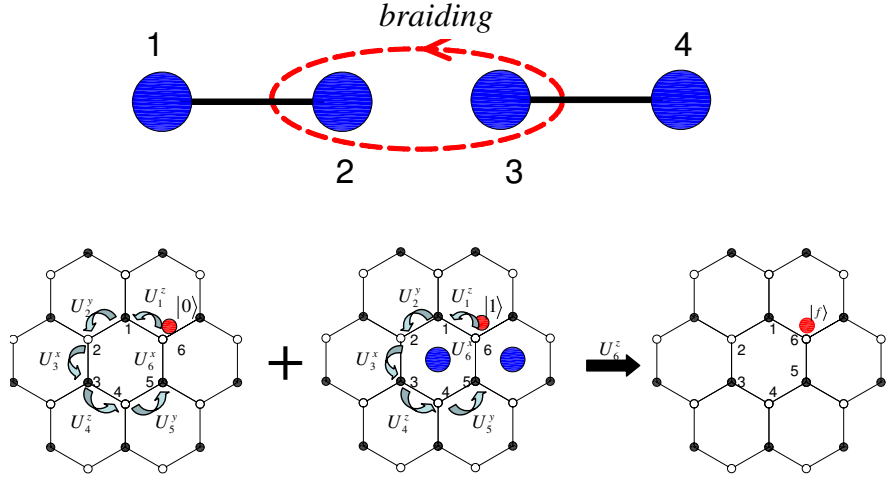}
\end{center}
\caption{(color online) The braiding of two non-Abelian anyons.
In a topological sense, a counterclockwise braiding (or exchange) of two anyons
is equivalent to move a anyon counterclockwise by $2\pi$ around the other.
We choose the braiding take place between anyons 2 and 3 due to superselection rule.
The operator $S_{23}=\sigma_5^y\sigma_4^z\sigma_3^x\sigma_2^y\sigma_1^z\sigma_6^x$
correspond to a looping trajectory which  is exactly the loop when  anyon 3 move around anyon 2.}
\end{figure*}

We next consider the simplest  vortices configuration that can experimentally reveal
the non-Abelian statistics as shown in Fig. 3, which consists of two pairs of vortices.
To reveal the non-Abelian statistics, we would like to move one anyon (Majorana fermion 3)
around the other (Majorana fermion 2), where  the two come from different Majorana fermion pairs.
Indeed, this braiding  can be implemented by the operator
$s_{23}=\sigma_5^y\sigma_4^z\sigma_3^x\sigma_2^y\sigma_1^z\sigma_6^x$ \cite{Kitaev2}.
For a system with 4 vortices, the degeneracy of the ground states is 4.
For an  adiabatic braiding of vortices 2 and 3 once, the  effect  is a unitary operation
acting on the the subspace of the ground states, which can be represented as \cite{Ivanov}
\begin{equation}
R_2=\frac{1}{\sqrt2}
\begin{pmatrix}
1 & -i \\ -i & 1
\end{pmatrix},
\end{equation}
where $R_i$ correspond to anticlockwise exchange between two Majorana fermions $i$ and $i+1$, and $R^{-1}$ represent a clockwise exchange.

For our particular implementation in Kitaev's honeycomb model, we prepare the initial state
of the ancilla atom in $|+\rangle_a \propto |0\rangle_a+|1\rangle_a$. We first create the
lattice atoms in the 2-vortex state $|v_2\rangle$, i.e., with anyons 1 and 2. Secondly,
bringing the ancilla close enough to the site "6" atom so that their collision interaction
is in effect, ie., implement a $U_z$ gate to the atom and the ancilla. Therefore,
the entangled state $|0\rangle_a|v_2\rangle+|1\rangle_a|v_4\rangle$  is produced with
$|v_4\rangle$ being the 4-vortex configuration state. By doing so, we have created anyons 3
and 4 on the condition of the ancilla in the state $|1\rangle_a$. Because  the braiding
will be implemented between  anyons 2 and 3 and the anyons 1 and 4 are not surrounded in
the braiding pathway, so it is convenient to work in the subspace of vortex 2 and 3 on
which operation $R_2$ act. Thirdly, move the ancilla close to the target atom "6" again
to realize $U_6^x$ gate operation, and then "1" to "5" successively  to realize the braiding
operation $S_{23}=U_5^yU_4^zU_3^xU_2^yU_1^zU_6^x$ when the
ancilla atom is in the $|1\rangle_a$ state (cf. Fig. 3 ).
After braiding of anyons $2$ and $3$ twice, we get
$|0\rangle_a|v_2\rangle+e^{-i\frac{\pi}{2}}|1\rangle_a R_2^2 |v_4\rangle$ with $R_2^2|v_4\rangle$
is a degenerate ground state with a fermionic occupation. The vortex pair state of 2 and 3
changes from the vacuum to a fermionic state with a fermion occupation.
Finally, repeat the second and first steps  to bring the system back to the vortex-free state,
the final state of the system is $(|0\rangle_a+e^{-i\frac{\pi}{2}}|1\rangle_a)|gs\rangle$
with the ancilla in the state of $|f\rangle \propto|0\rangle_a+e^{-i\frac{\pi}{2}}|1\rangle_a$.
Then, compare the initial  and the final states of the ancilla one can find that a phase factor
$-\pi/2$ is generated due to the non-Abelian nature of anyons. The recorded interferometry result
of the non-Abelian statistics can be read from the ancilla atom by a local projective measurement
after a $\pi/4$ phase gate. The single atom gate and measurement can be realized with very high efficiency.

Note that while the anyon obeys Abelian statistics, our procedure will bring
the final state of the ancilla to $|0\rangle-|1\rangle$ (braiding once);
for  bosonic or fermionic statistics, the ancilla's  state will stay unchanged.
Therefore, probe the final state of the ancilla will verify the no-Abelian statistics.
On the other hand, for no-Abelian anyons, when the  braiding is repeated  twice and four times,
the final states of the ancilla atom will be the same as the Abelian anyon case (braiding once)
and be identical to the initial state (up to an overall $\pi$ phase as $R_2^4=-1$),
respectively. For  Abelian  anyon, two successive identical braiding will lead
the ancilla's final state to be the same as its initial state.  This can serve as
further distinction of the  Abelian and no-Abelian anyons in this particular model.

In summary, we propose a scheme to probe the non-Abelian statistics of  Majorana
fermions bounding to the vortex excitation in the Kitaev honeycomb lattice model
with effective magnetic field using cold atom in an optical lattice. The exotic non-Abelian
statistics can be  recorded and detected by the internal state of the ancilla atom,
the measurement of which can be realized with  very high efficiency by standard techniques.

\section*{Acknowledgments}

This work was supported by the  NSFC (No. 11004065), the NFRPC (No. 2013CB921804),
and the PCSIRT.

\end{document}